\PassOptionsToPackage{numbers,sort,compress}{natbib}

\documentclass{article} 
\usepackage{iclr2026_conference,times}

\usepackage[inline]{enumitem}
\usepackage[utf8]{inputenc} 
\usepackage[T1]{fontenc}    
\usepackage{hyperref}       
\usepackage{url}            
\usepackage{booktabs}       
\usepackage{amsfonts}       
\usepackage{nicefrac}       
\usepackage{microtype}      
\usepackage[skip=2pt]{caption}
\usepackage{wrapfig}
\usepackage{graphicx} 
\usepackage{xcolor}         
\usepackage{booktabs}
\usepackage{graphicx} 
\usepackage{mathtools}
\usepackage{wrapfig}
\usepackage{algorithm}
\usepackage{algorithmic}
\usepackage{booktabs,siunitx,adjustbox}

\definecolor{darkblue}{rgb}{0, 0, 0.5}
\hypersetup{colorlinks=true, citecolor=darkblue, linkcolor=darkblue, urlcolor=darkblue}

\newcommand{\edit}[1]{\textcolor{black}{#1}}

\usepackage{colortbl,xcolor}
\definecolor{heatblueMin}{RGB}{173,210,233}   
\definecolor{heatblueMax}{RGB}{255,255,255}   
\definecolor{heatorangeMin}{RGB}{255,224,178} 
\definecolor{heatorangeMax}{RGB}{255,255,255} 

\newcommand{\heatblue}[2]{\cellcolor{heatblueMin!#1!heatblueMax}{#2}}
\newcommand{\heatorange}[2]{\cellcolor{heatorangeMin!#1!heatorangeMax}{#2}}

\usepackage{listings}
\usepackage{xcolor}

\usepackage{mdframed}

\title{ZeroGR: A Generalizable and Scalable Framework for Zero-Shot Generative Retrieval}

\author{Weiwei Sun\textsuperscript{\rm 1, $*$}\quad
    Keyi Kong\textsuperscript{\rm 2, }\thanks{Equal contribution.}\quad
    Xinyu Ma\textsuperscript{\rm 3}\quad
    Shuaiqiang Wang\textsuperscript{\rm 3} \\
    \textbf{Dawei Yin\textsuperscript{\rm 3} \quad
    Maarten de Rijke\textsuperscript{\rm 4}\quad
    Zhaochun Ren\textsuperscript{\rm 5, }\thanks{Corresponding author.}\quad
    Yiming Yang\textsuperscript{\rm 1}} \\
    $^{1}$Carnegie Mellon University \quad
    $^{2}$Shandong University \quad  
    $^{3}$Baidu Inc \\
    $^{4}$University of Amsterdam \quad 
    $^{5}$Leiden University \\
    \texttt{\{sunnweiwei,luxinyayaya012,xinyuma2016.com,shqiang.wang\}@gmail.com}\\
    \texttt{yindawei@acm.org}, \texttt{m.derijke@uva.nl},
    \texttt{z.ren@liacs.leidenuniv.nl}, \\\texttt{yiming@cs.cmu.edu}
}
\iclrfinalcopy

\begin{document}

\maketitle

\begin{abstract}
Generative retrieval (GR) reformulates information retrieval (IR) by framing it as the generation of document identifiers (docids), thereby enabling end-to-end optimization and seamless integration with generative language models (LMs).
Despite notable progress under supervised training, GR still struggles to generalize to zero-shot IR scenarios, which are prevalent in real-world applications. 
To tackle this challenge, we propose \textsc{ZeroGR}, a zero-shot generative retrieval framework that uses natural language instructions to extend GR across a wide range of IR tasks.
Specifically, \textsc{ZeroGR} is composed of three key components: (i) an LM-based docid generator that unifies heterogeneous documents (e.g., text, tables, code) into semantically meaningful docids; (ii) an instruction-tuned query generator that generates diverse types of queries from natural language task descriptions to enhance corpus indexing; and (iii) a reverse annealing decoding strategy to balance precision and recall during docid generation. 
Furthermore, we introduce OpenInstIR, the most diverse open-source instructed retrieval dataset.
We investigate the impact of instruction fine-tuning scale and find that performance consistently improves as the number of IR tasks encountered during training increases.
Extensive experiments on the BEIR and MAIR benchmarks demonstrate that \textsc{ZeroGR} achieves competitive performance across a wide range of retrieval tasks, establishing a new state-of-the-art among GR methods. Our code is available at \url{https://github.com/sunnweiwei/ZeroGR}.
\end{abstract}

\section{Introduction}

Dense retrieval (DR) is arguably the most effective and widely adopted information retrieval (IR) paradigm today~\citep{Karpukhin2020DensePR,Izacard2021UnsupervisedDI,Thakur2021BEIRAH,Muennighoff2022MTEBMT}. It encodes documents and queries as embedding vectors.
Despite its success, DR's expressivity is fundamentally limited by the embedding dimensionality~\citep{DeCao2020AutoregressiveER} and does not fully use the capabilities of generative language models (LMs)~\citep{Tay2022TransformerMA}.
As an alternative, generative retrieval (GR)~\citep{Metzler2021RethinkingSM} introduces a paradigm shift that encodes corpus information into the model parameters, enabling document retrieval  by generating (relevant) document identifiers (docids).
GR has demonstrated competitive performance on various IR tasks when large-scale supervised data is available~\citep{Tay2022TransformerMA,Sun2023LearningTT,Chen2022CorpusBrainPA}, spanning both traditional web search~\citep{Campos2016MSMA} and knowledge-intensive retrieval applications~\citep{Petroni2020KILTAB}.

Despite its promising performance on in-domain tasks, GR still exhibits limited generalization to out-of-distribution IR tasks.  
Existing GR models are typically trained on specific corpora and queries, and prior studies have shown that such training leads to poor performance on unseen tasks~\citep{zhang2025replication,liu2023robustness}.
In contrast, real-world IR models are typically evaluated in a broader setting, characterized by substantial diversity and heterogeneity.
These often involve heterogeneous corpora and queries~\citep{Thakur2021BEIRAH}, task-specific relevance criteria~\citep{Su2022OneEA,Asai2022TaskawareRW}, and predominantly zero-shot scenarios where no supervised data is available~\citep{Thakur2021BEIRAH,Muennighoff2022MTEBMT}. 
Consequently, GR approaches designed for supervised conditions struggle to generalize to such heterogeneous and data-scarce retrieval scenarios.

To address the limitations of GR in zero-shot and heterogeneous IR scenarios, we
propose \textbf{\textsc{ZeroGR}}, a generalizable framework for \textbf{ZERO}-shot \textbf{G}enerative information \textbf{R}etrieval. \textsc{ZeroGR} is a simple yet effective way to adapt GR to diverse IR tasks in a zero-shot setting by using natural language task instructions.
Specifically, we advance GR along three dimensions:
(i)~for \textit{docid design}, we propose a docid generator to efficiently convert a document of any format (e.g., paragraph, table, code) into a unified text-based docid representation;
(ii)~for \textit{corpus indexing}, we propose an instructed query generator to generate diverse types of queries based on different task instructions; and
(iii)~for \textit{docid decoding}, we propose a reverse annealing strategy that more effectively  trades off precision and recall of docid decoding than prior work. 

\begin{figure}[t]
    \centering
        \includegraphics[width=1\textwidth]{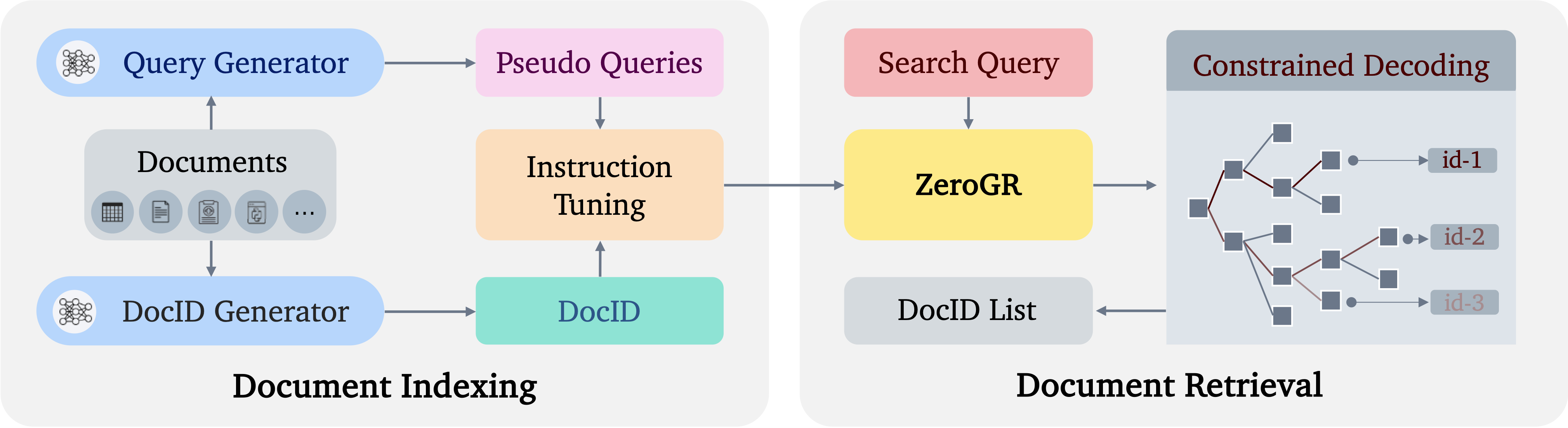}
    \vspace*{-3mm}
    \caption{An overview of \textsc{ZeroGR}. Given a document collection, \textsc{ZeroGR} converts documents in the collection into unified DocID representations, generates diverse pseudo-queries, and builds a generative retrieval index. During online retrieval, \textsc{ZeroGR} decodes docids with reverse-annealed temperature scheduling to balance precision and recall.}
    \label{fig:diagram}
\end{figure}

Building on \textsc{ZeroGR}, we investigate instruction fine-tuning along two key axes: the size of instruction tuning data and the size of the underlying model.
Existing studies typically rely on single-domain~\citep{Weller2024PromptrieverIR} or closed-source training data~\citep{Muennighoff2024GenerativeRI}, hindering a systematic analysis of generalization. Instead, we collect the first large-scale open-source retrieval instruction dataset \textbf{OpenInstIR} (Open Instructed
Information Retrieval Dataset), spanning diverse domains and tasks, enabling rigorous study of instruction generalizability in zero-shot retrieval.
Through extensive experiment based on OpenInstIR, we find that increasing both the diversity and quantity of training tasks yields substantial improvements in zero-shot retrieval performance on unseen tasks.
Beyond training data scaling, we also examine model size scaling and inference-time scaling for corpus indexing, observing consistently promising scaling trends in both cases.
\edit{Our best-performing model, based on Llama-3B, outperforms previous generative retrieval methods and narrows the gap to state-of-the-art dense retrieval systems across heterogeneous IR benchmarks~\citep{Thakur2021BEIRAH,Sun2024MAIRAM}}.
Notably, \textsc{ZeroGR} performs on par with OpenAI Embed-v3 on zero-shot MAIR tasks, highlighting its strong generalization to unseen retrieval tasks.

In summary, our contributions are as follows:
\begin{enumerate}
    \item We propose \textsc{ZeroGR}, a zero-shot GR framework that can construct task-specific GR search indices based on natural language instructions.
    \item Within \textsc{ZeroGR}, we enhance GR by introducing three key components: a unified text-based docid generator, an instruction-conditioned pseudo-query generator, and a reverse annealing decoding strategy.
    \item We collect the first large-scale open-source instruction-tuning dataset for retrieval, OpenInstIR, that spans diverse domains and task formulations, enabling systematic study of model generalization in retrieval.
    \item \textsc{ZeroGR} achieves competitive performance on heterogeneous IR benchmarks, establishing it as the first GR approach capable of generalizing to diverse tasks in a zero-shot setting.
\end{enumerate}

\section{Related Work}

\paragraph{Document retrieval.}
Document retrieval is a fundamental task in information retrieval, with broad applications in search engines and retrieval-augmented generation systems~\citep{Karpukhin2020DensePR,Lin2020PretrainedTF,Chen2025ImprovingRG}.
It typically follows a two-stage pipeline: an initial retrieval stage that recalls candidate documents, followed by a reranking stage for fine-grained ranking.
Traditional sparse retrieval methods~\citep{Robertson1997OnRW,Lafferty2001DocumentLM,Robertson2009ThePR} rely on lexical overlap but suffer from vocabulary mismatch~\citep{Lin2020PretrainedTF}.
Dense retrieval (DR) addresses this issue by embedding queries and documents into dense vectors and comparing them via inner product or cosine similarity~\citep{Karpukhin2020DensePR}, with subsequent improvements from hard negative mining, late interaction, and pre-training~\citep{Xiong2020ApproximateNN,Khattab2020ColBERTEA,Wang2022TextEB,Qu2020RocketQAAO,Izacard2021UnsupervisedDI}.
The reranking stage is usually performed using cross-encoders or LLM prompting~\citep{Nogueira2019PassageRW,Nogueira2020DocumentRW,Sun2023IsCG,Chen2024TourRankUL,Zhang2025Qwen3EA,Liu2025ReasonRankEP,Ma2023FineTuningLF}.

However, this two-stage pipeline is difficult to optimize end-to-end due to its MIPS-based retrieval component and the objective mismatch with generative language model training~\citep{Tay2022TransformerMA,Bevilacqua2022AutoregressiveSE}.

\paragraph{Generative retrieval.}
Unlike traditional dense retrieval methods~\citep{Karpukhin2020DensePR,Xiong2020ApproximateNN}, GR formulates information retrieval as a docid generation task, enabling end-to-end optimization of the inference-time search index~\citep{Tay2022TransformerMA,Metzler2021RethinkingSM}.
Previous research on GR has largely focused on three key aspects:
(i) \textit{Docid design}: Early approaches employed rule-based formats such as titles~\citep{DeCao2020AutoregressiveER,Chen2022CorpusBrainPA}, URLs~\citep{Zhou2022UltronAU}, or text spans/summaries~\citep{Bevilacqua2022AutoregressiveSE,Li2023SummarizationBasedDI}. More recent work has shifted toward learning-based docid designs that capture corpus semantics more effectively, including embedding clustering~\citep{Tay2022TransformerMA} and RQ-VAE–based approaches~\citep{Wang2024ContentBasedCG,Zeng2023ScalableAE,Wang2023NOVOLA}. 
(ii) \textit{Corpus indexing}: Several strategies have been explored to enrich corpus representations, such as document chunking~\citep{Tay2022TransformerMA}, pseudo-query generation~\citep{Zhuang2022BridgingTG}, rehearsal-based augmentation~\citep{Tang2023SemanticEnhancedDS}, multi-granular indexing~\citep{Wen2025OnSD}, and continual training for dynamic corpora~\citep{Mehta2022DSIUT,Chen2023ContinualLF,zhang2025replication}.  
(iii) \textit{Docid decoding}: The dominant approach has been constrained beam search~\citep{DeCao2020AutoregressiveER,Tay2022TransformerMA}. More advanced strategies include multi-stage decoding~\citep{Ren2023TOMEAT}, multi-docid decoding~\citep{Li2023MultiviewIE}, and simultaneous decoding~\citep{Zeng2024PlanningAI}.  

Despite steady progress, existing work primarily remains confined to supervised fine-tuning, relying heavily on training data and failing to generalize to zero-shot retrieval tasks. 

\paragraph{Instruction fine-tuning in IR.}
Inspired by studies in LLM instruction tuning~\citep{Chung2022ScalingIL,Wang2022SuperNaturalInstructionsGV}, instruction fine-tuning for retrieval has gained increased attention to improve zero-shot IR performance~\citep{Su2022OneEA,Asai2022TaskawareRW}.
Instruction-tuned models are able to adapt to various tasks based on natural language instructions that specify the relevance criteria.
Recent studies in this direction include multi-task fine-tuning~\citep{Lee2024NVEmbedIT}, LLM-generated instruction data~\citep{Wang2023ImprovingTE,Lee2024GeckoVT,Oh2024INSTRUCTIRAB}, and instruction-negatives~\citep{Weller2024PromptrieverIR}.
These efforts have primarily focused on dense retrieval or cross-encoder rerankers~\citep{Sun2024MAIRAM}.
To the best of our knowledge, we are the first to investigate instruction fine-tuning for GR and to conduct a systematic study of the factors that influence instruction fine-tuning.

\section{Preliminaries}

\textbf{Zero-shot document retrieval.}
We formulate the task of zero-shot document retrieval as follows. 
Given a corpus $\mathcal{D} = (d_1, \ldots, d_n)$ containing $n$ documents, a \textit{corpus indexing} function $\mathcal{I}$ takes $\mathcal{D}$ as input and constructs a search index $m = \mathcal{I}(\mathcal{D})$. 
Then, a \textit{retrieval} function $\mathcal{F}$ takes the index $m$ and a query $q$ as input, and returns a list of relevant documents: $(d_i, \ldots) = \mathcal{F}(m, q)$.
Note that in a typical zero-shot document retrieval setting, no training data is available.
However, a natural language task instruction $\textit{$instr_t$}$ specifying the retrieval task is generally assumed to be available, as it is usually easier to obtain~\citep{Muennighoff2022MTEBMT}.

In dense retrieval~\citep{Karpukhin2020DensePR}, the indexing function can be defined as using a document encoder to encode the corpus as a embedding matrix such as $\mathbf{E} \in \mathbb{R}^{n \times k}$ and the index is defined as the matrix $m \coloneqq \mathbf{E}$.
Then for retrieval function, a query encoder encode the query $q$ as $\mathbf{q}\in \mathbb{R}^{1\times k}$ and then perform maximum inner-product search (MIPS) over index $m$ to find closest document in embedding space.

\textbf{Generative retrieval.}
GR aims to retrieve the document $d_i$ by generating the corresponding document identifier (docid) given the query $q$. 
To this end, GR assigns an identifier (docid) to each document in the corpus, e.g. $(z_1, \ldots, z_n)$, where each $z_i$ is a sequence of tokens $z_i = \{z_i^{(1)}, \ldots, z_i^{(T)}\}$ with a maximum length of $T$.
Based on this, the indexing function $\mathcal{I}(\mathcal{D})$ of GR is to train a language model (LM) $\mathcal{M}$ on the corpus $\mathcal{D}$, encoding the corpus information and also document-docid mapping.
The retrieval function $F$ is instantiated by the same $\mathcal{M}$, and it generates the relevant document identifiers (docids) $(z_1, \ldots, z_n)$ given the query $q$: $(z_i, \ldots) = \mathcal{M}(q)$.

\section{\textsc{ZeroGR}}
We propose \textsc{ZeroGR}, a zero-shot GR framework that can adapt LMs into task-specific generative search indexes based on task instructions.
As shown in Figure~\ref{fig:diagram}, the \textsc{ZeroGR} framework consists of three key components:
(i) a docid generator $G_{\psi}$, which takes a document $d_i$ as input and outputs its docid $z_i$;
(ii) an instructed query generator, which takes a task instruction  $\textit{instr}$ and a document $d_i$ as input and outputs multiple pseudo-queries; and
(iii) a generative retriever $\mathcal{M}$, which takes the instruction and a query as input and generates a list of docids.

The \textsc{ZeroGR} pipeline proceeds as follows:
(i)~given a new corpus $\mathcal{D}$ and its associated task instruction $\textit{instr}$, the docid generator assigns each document $d_i$ a docid $z_i$;
(ii)~the instructed query generator $G_{\theta}$ samples $B$ queries $\{q_{i,1}, \ldots, q_{i,B}\}$ for each document $d_i \in \mathcal{D}$, thereby creating $\langle q_{i,j}, z_i \rangle$ pairs; and
(iii)~the generative retriever is trained to predict the corresponding docid $z_i$ given the concatenation of $\textit{instr}$ and a sampled query $q_{i,j}$.
After training, the generative retriever $\mathcal{M}(z \mid q, \textit{instr})$ serves as the search index $m$. For a given query $q$, a newly proposed reverse annealing decoding strategy is employed to generate a ranked list of docids as retrieval results.

\subsection{Unified docid representation}
Documents in downstream IR tasks can be heterogeneous, e.g., financial tables~\citep{Zhu2022TowardsCD}, code files~\citep{Liu2023RepoBenchBR}, meeting transcripts~\citep{Golany2024EfficientDG}, or legal cases~\citep{Bhattacharya2019OverviewOT}.
Existing simple docid strategies, such as using document titles, URLs, or spans~\citep{DeCao2020AutoregressiveER,Bevilacqua2022AutoregressiveSE}, often fail to generalize to user-customized data.
\textsc{ZeroGR} therefore introduces a model-based \textbf{docid generator} \(G_{\psi}\) that maps any document to a short, keyword-rich sentence (typically 6–8 words) ranked by coverage.  
Formally, for a document $d_i$ we define the docid $z_i$ as follows:
\begin{equation}
\label{eq:docid}
z_{i}
\;=\;
G_{\psi}(d_i)
\;=\;
\underset{t\in\mathcal{V}^{\le L}}{\arg\max}\,
G_{\psi}\!\bigl(t \mid d_i\bigr),
\end{equation}
where \(t\) is a token sequence of length \(\le L\) (with \(L=8\)) drawn from the vocabulary~\(\mathcal{V}\).
To instantiate \(G_{\psi}\), we first prompt a powerful LM (e.g., GPT-4o) to create a training set of \(\langle d_i,\,z_i\bigr\rangle\) pairs. 
A smaller model (Llama-3.2-1B) is then fine-tuned on this data, enabling fast, scalable generation of unified docids across diverse IR tasks. Section~\ref{sec:train_data} details our training data.

\subsection{Instructed Corpus Indexing}

Corpus indexing in GR encodes each document \(d_i\in\mathcal{D}\) into the model's parameters so that, at inference time, the model can recover $d_i$ by \emph{generating} its document identifier \(z_i\).  
DSI-QG~\citep{Zhuang2022BridgingTG} accomplishes this by pairing every document with a set of pseudo-queries, but its effectiveness diminishes when the pseudo-query distribution diverges from real user queries~\citep{Pradeep2023HowDG,Dai2022PromptagatorFD}.  
This gap is especially large in heterogeneous IR scenarios, such as conversational, code, or multimodal search.

We mitigate the distribution gap with an \textbf{instructed query generator} \(G_{\theta}\), obtained by instruction-tuning a 1B-parameter Llama model on diverse IR datasets verbalized through task-specific instructions.  
Given a document $d_i$ and a task instruction \(\textit{instr}\), the generator produces a pseudo-query $q_{i,j}$ from the conditional distribution
\begin{equation}
\label{eq:generator}
q_{i,j} \;\sim\; G_{\theta}\bigl(\cdot \mid d,\textit{instr}\bigr).
\end{equation}
For each document we draw $B$ queries with a temperature of 1:
\begin{equation}
\mathcal{Q}_{i}=\{\,q_{i,1},\dots,q_{i,B}\,\}.
\end{equation}
These \(\langle d_i,\,z_i\bigr\rangle\) pairs are used to train the generative retriever $\mathcal{M}$ by minimizing the cross-entropy loss
\begin{equation}
\label{eq:ce-loss}
\mathcal{L}(\phi)
=
-\!\!\!\sum_{d_i\in\mathcal{D}}\;\sum_{q_{i,j}\in\mathcal{Q}_{i}}
\log \mathcal{M}\bigl(z_{i}\mid q_{i,j},\textit{instr}\bigr),
\end{equation}
thereby embedding the corpus into the model’s parameters.
Table~\ref{tab:domain_stats} summarizes the instruction-tuning datasets.

\subsection{Reverse-annealed docid generation}
During inference, a GR model must decode each docid \(z_{i}\) as a \emph{sequence of tokens}.  
Standard beam search often collapses to a few high-probability sequences, hurting recall~\citep{Wu2025ConstrainedAD}.  
We therefore propose \textbf{reverse-annealed sampling}: each \(z_{i}\) is generated token-by-token, while the sampling temperature is gradually \emph{increased} to encourage diversity.
Let \(f(\cdot)\) denote the trained decoder after corpus indexing, and let \(T\) be a prefix tree whose leaves correspond to valid docids.  
For the \(i\)-th docid we decode a token sequence \(\mathbf{x}_{i}=(x_{i,1},\dots,x_{i,L_{i}})\) using temperature \(t_{i}=g(i)\).  
At position \(j\) we sample
$
x_{i,j}\;\sim\;
\operatorname{Softmax}\Bigl(\tfrac{\boldsymbol{\ell}_{i,j}}{t_{i}}\Bigr)\Big|_{T_{i,j}},
$
where \(\boldsymbol{\ell}_{i,j}\) are the logits conditioned on the current prefix
\((x_{i,1{:}j-1})\), and the subscript \(T_{i,j}\) masks probabilities to tokens that keep the prefix inside the tree.
After the complete sequence \(\mathbf{x}_{i}\) is produced, its leaf is removed from \(T\) so no subsequent iteration can repeat the same docid.
The per-iteration temperature \(t_{i}\) follows a \emph{normalized sigmoid}:
\begin{equation}\small
\label{eq:temperature}
t_{i}
=
g(i)
=
T_{\max}\cdot
\frac{\sigma\!\bigl(k(\tfrac{i}{K}-m)\bigr)-\sigma(-km)}
     {\sigma\!\bigl(k(1-m)\bigr)-\sigma(-km)},
\quad
\sigma(z)=\frac{1}{1+e^{-z}},
\end{equation}
where \(K\) is the total number of docids to generate, \(k>0\) controls the slope, and \(m\in(0,1)\) sets the midpoint.  
Starting from a low temperature yields high-precision early selections; increasing \(t_{i}\) over iterations boosts exploration, thereby balancing precision and recall across the final ranked list. \edit{See Algorithm~\ref{alg:reverse_annealing} and Figure~\ref{fig:normalized-sigmoid}, both in the Appendix, for details.}









\section{Experimental Design}

Our experiments address the following research questions:

\begin{enumerate}
    \item \textbf{How do model design and training strategies influence the performance of \textsc{ZeroGR}?} \quad
    To answer this, we conduct a systematic study on the development set, investigating key factors in generative retrieval. Specifically, we analyze how instruction tuning task diversity (Section~\ref{sec:task_num}), docid design (Section~\ref{sec:docid}), corpus indexing strategy, decoding strategy (Section~\ref{sec:decode}), and model size (Section~\ref{sec:query_num}) affect performance.
    \item \textbf{How does \textsc{ZeroGR} compare with dense retrieval methods?} \quad
    We evaluate \textsc{ZeroGR} against leading models on the MAIR benchmark (Section~\ref{sec:mair}) and conduct additional analysis on the BEIR datasets (Section~\ref{sec:beir}).
    
\end{enumerate}

\subsection{OpenInstIR: Open Instructed Information Retrieval Dataset}
\label{sec:train_data}
Existing studies typically rely on single-domain~\citep{Weller2024PromptrieverIR} or closed-source training data~\citep{Muennighoff2024GenerativeRI}.
To support the development of \textsc{ZeroGR}, we collect training data covering a diverse range of IR tasks.
Specifically, we extract the training splits of public information retrieval datasets (e.g., those in \citep{Sun2024MAIRAM,Muennighoff2024GenerativeRI}) and construct a multi-task training set with annotated task instructions and relevance labels.

As shown in Table~\ref{tab:data_stats}, \textbf{OpenInstIR} spans 69 IR tasks across 6 domains and contains 41 million query-document pairs.
OpenInstIR is the largest open-source IR training corpus to date, offering greater domain and task diversity, detailed instructional annotations, and reliable relevance labels.
See Appendix~\ref{sec:train-data} for detailed statistics.

\begin{wraptable}[13]{r}{.5\textwidth}
  \centering\small
\setlength{\tabcolsep}{1pt}
\vspace*{-\baselineskip}
\caption{\edit{Dataset domain statistics.}}
\label{tab:data_stats}
\begin{tabular}{lcrcrcr}
\toprule
 & \multicolumn{2}{c}{\textbf{OpenInstIR}}
 & \multicolumn{2}{c}{\textbf{MAIR-Test}}
 & \multicolumn{2}{c}{\textbf{BEIR}} \\
\cmidrule(r){2-3}
\cmidrule(r){4-5}
\cmidrule(r){6-7}
\textbf{Domain}
 & Tasks & Size
 & Tasks & Size
 & Tasks & Size \\
\midrule
Academic & 18 & 744,160 & 5 & 500 & 2 & 200 \\
Code     & 13 & 1,969,586 & 3 & 300 & 0 & 0 \\
Finance  & 8  & 31,315 & 5 & 439 & 1 & 100 \\
Legal    & 7  & 23,086,948 & 4 & 300 & 0 & 0 \\
Medical  & 5  & 421,430 & 8 & 459 & 2 & 150 \\
Web      & 18 & 15,319,445 & 13 & 935 & 6 & 569 \\
\midrule
All      & 69 & 41,572,884 & 38 & 2,933 & 11 & 1,019 \\
\bottomrule
\end{tabular}
\end{wraptable}

\subsection{Evaluation datasets}
To evaluate zero-shot GR on diverse downstream tasks, we use the BEIR and MAIR benchmarks:
(i)~\textbf{BEIR}~\citep{Thakur2021BEIRAH}. We evaluate models on all 12 tasks from the BEIR collection.
(ii)~\textbf{MAIR}~\citep{Sun2024MAIRAM}. As we collect training data from a subset of MAIR tasks, we divide MAIR into seen and unseen subsets, where the unseen subset contains tasks not present in the OpenInstIR training data, to validate the zero-shot generalization of models. 
In constructing this benchmark, we curated a diverse set of long-tail tasks across 6 domains, and intentionally omitted redundant tasks (e.g., different years of the same competition) and structurally complex ones (e.g., IFEval) that would introduce evaluation overhead.
Given the large size of the MAIR dataset, we also develop a Dev subset of MAIR for model ablation (see the detailed task breakdown in Table~\ref{tab:dev-set}).
\edit{Note that our current evaluation focuses on tasks with moderately sized corpora.}

\subsection{Evaluation metric}
We evaluate models using the following metrics:
(i)~\textit{Top-1 accuracy}, which measures retrieval precision by checking whether the top-ranked document is relevant to the query;
(ii)~\textit{nDCG@10}, evaluates the quality of the top-10 ranked results by considering both the relevance and position of retrieved documents; and
(iii)~\textit{Recall@100}, which assesses recall by calculating the percentage of relevant documents retrieved within the top-100 ranked list.

\subsection{Implementation details}
We implement the three components of \textsc{ZeroGR}, i.e., query generator, docid generator, and final generative retriever, all with Llama-based LMs.
For the docid generator, a Llama-1B-Instruct model is trained on our curated document-docid pairs for 5 epochs with a constant learning rate of 5e-5.
Similarly, for the query generator, a Llama-1B-Instruct model is trained on the OpenInstIR training set for 5 epochs with a constant learning rate of 5e-5.
For the generative retriever, the model is trained for each evaluated task on data generated by the query generator and docid generator, 
based on our ``Document Indexing'' workflow described in Figure~\ref{fig:diagram}.

\subsection{Baselines}


We evaluate \textsc{ZeroGR} against several representative IR baselines, spanning different retrieval paradigms to provide a comprehensive comparison. 
\begin{enumerate}
\item For sparse retrieval, we adopt the classical term-based model \textbf{BM25}, implemented using the BM25S package~\citep{L2024BM25SOO}.

\item For traditional dense retrieval models trained on a single task, we include \textbf{Contriever-MARCO}, \textbf{GTR-base}, and \textbf{GTR-Large}, all of which are pretrained or fine-tuned on the MS MARCO dataset~\citep{Ni2021LargeDE, Izacard2021UnsupervisedDI}, representing a common practice in dense retrieval pipelines. 

\item For multi-task-trained dense retrievers, we incorporate \textbf{E5-Base} and \textbf{E5-Large}~\citep{Wang2022TextEB}, \textbf{BGE-base} and \textbf{BGE-Large}~\citep{Xiao2023CPackPR}, as well as \textbf{OpenAI-Embedding-v3-Small}, all of which use supervision from multiple tasks to enhance generalization across diverse domains. 

\item For instruction-tuned dense retrieval models, which aim to align the retriever with human instructions, we include \textbf{E5-Mistral-7B-instruct}~\citep{Wang2023ImprovingTE}, and \textbf{GritLM-7B}~\citep{Muennighoff2024GenerativeRI}, which are trained on large-scale, diverse instruction datasets to follow task-specific intents effectively. 
\end{enumerate}

\begin{figure*}[t]
  \centering
  \includegraphics[width=0.9\textwidth]{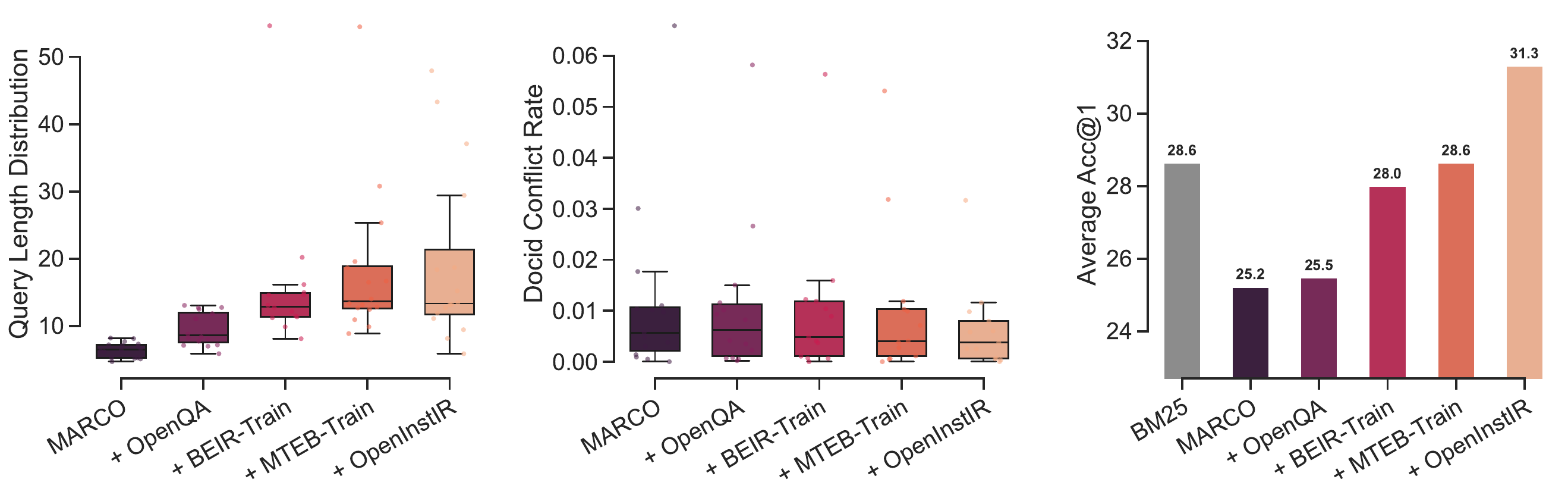}
  \caption{Model performance on unseen-dev tasks as a function of the number of training tasks.
  We increase the number of training tasks, starting from MS MARCO, and incrementally add open-domain QA datasets (e.g., NQ), BEIR-Train sets (e.g., NFC), MTEB-Train data (e.g., NLI), and finally the OpenInstIR collection, which includes 60 tasks across 6 domains.
  (Left): More instruction-tuning tasks lead to more diverse queries.
  (Middle): More instruction-tuning tasks reduce docid conflicts.
  (Right): More instruction-tuning tasks improve the Acc@1 score.
  }
  \label{fig:task_num}
\end{figure*}

\section{Ablation Study}

We conduct a systematic study on the development set, analyzing how task diversity in instruction tuning (Section~\ref{sec:task_num}), docid design (Section~\ref{sec:docid}) and distribution (Section~\ref{sec:docid-dist}), corpus indexing strategy, decoding strategy (Section~\ref{sec:decode}), and model size (Section~\ref{sec:query_num}) affect generative retrieval performance.

\subsection{Analysis of task diversity}
\label{sec:task_num}

A key factor in enhancing the performance of LM-based tasks is scaling, i.e., increasing model size or data volume. 
The effectiveness of \textsc{ZeroGR} stems from instruction fine-tuning on multi-task IR datasets, which improves the instruction-following abilities of both the query generator and the title generator models.
To investigate the impact of multi-task training, we curate training data with varying numbers of tasks:  
(a) \textit{MS MARCO}, which contains a single task (i.e., MS MARCO~\citep{Campos2016MSMA}) and is commonly used in previous GR work;  
(b) \textit{+ OpenQA}, which adds popular open-domain question answering datasets, including NQ~\citep{Kwiatkowski2019NaturalQA} and HotpotQA;  
(c) \textit{+ BEIR-Train}, which incorporates the training splits of BEIR~\citep{Thakur2021BEIRAH}, such as NFCorpus and Quora;  
(d) \textit{+ MTEB-Train}, which includes additional tasks from MTEB~\citep{Muennighoff2022MTEBMT} that are not covered in BEIR, such as NLI (we use the public BGE training split to collect these data); and  
(e) \textit{+ OpenInstIR}, which includes the data we collected from the training split of the MAIR~\citep{Sun2024MAIRAM} task collection, comprising 69 tasks from 6 domains (Figure~\ref{fig:task_num}).

Figure~\ref{fig:task_num} shows the evaluation results of models (both query generator and docid generator) trained with different levels of task diversity, evaluated on the unseen task subset (i.e., tasks not included in any training set) of MAIR.
The left plot in Figure~\ref{fig:task_num} shows the distribution of average query length across tasks. We observe that models trained on more IR tasks generate queries with greater length diversity, indicating task-aware query generation strategies. In contrast, the baseline model trained only on MS MARCO produces short queries, averaging 8 words.
The middle plot shows the docid conflict rate, i.e., the percentage of documents in the corpus assigned the same docid by the docid generator. Models trained on diverse tasks exhibit lower conflict rates, suggesting a stronger ability to process heterogeneous corpora. The MS MARCO baseline shows higher conflict on several diverse tasks.
Finally, the right plot reports retrieval performance (top-1 accuracy) for different models. We observe consistent performance improvements on unseen tasks as training data diversity increases. 

\begin{figure}[t]
  \centering
  \includegraphics[width=.37\textwidth]{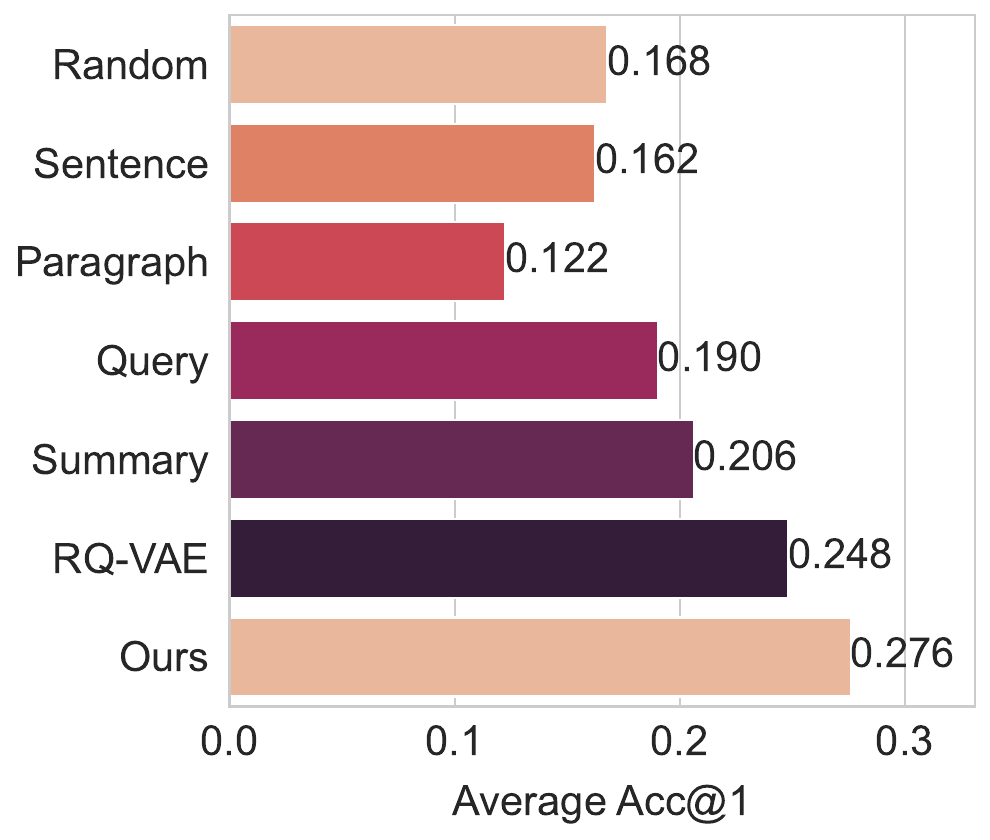}\hfill
  \includegraphics[width=.31\textwidth]{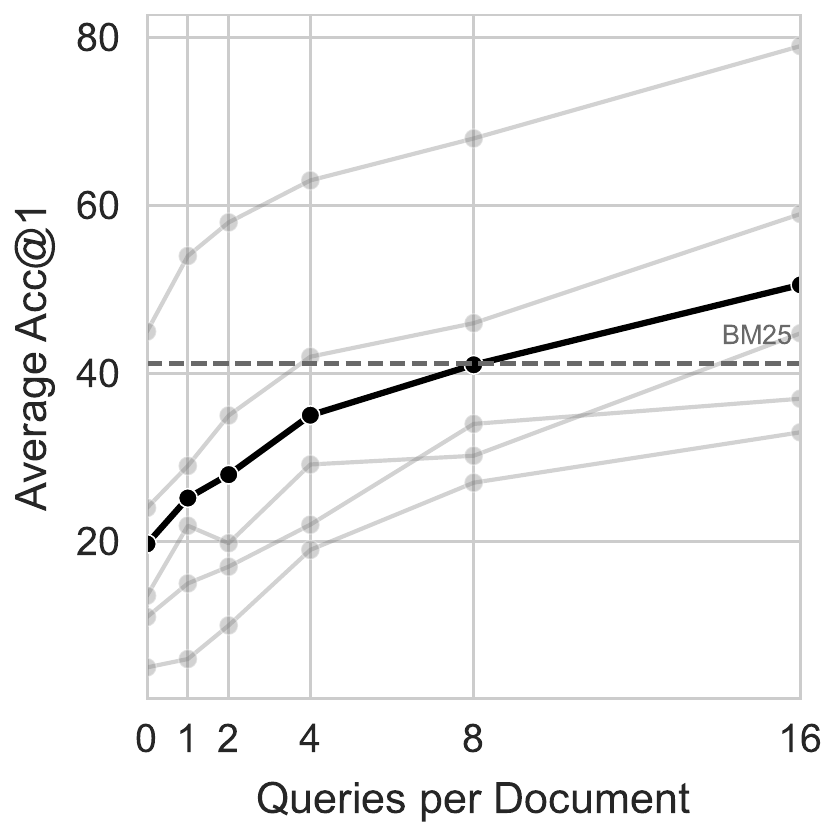}\hfill
  \includegraphics[width=.31\textwidth]{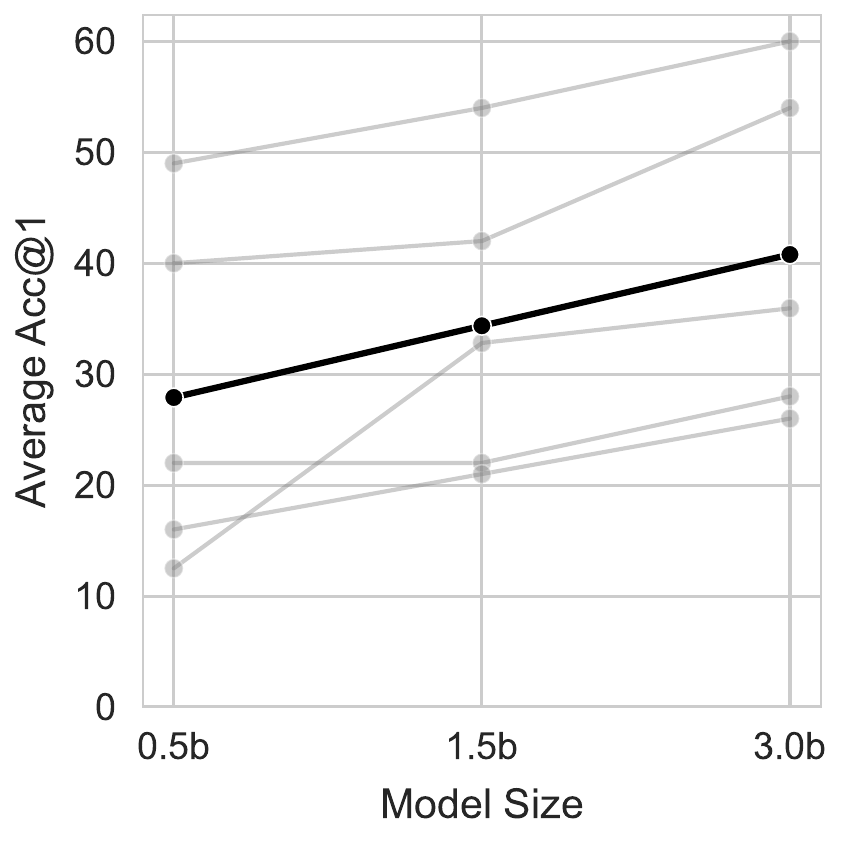}
  \caption{
  (Left): Comparison of different docid designs.
  (Middle): Acc@1 vs. generated queries per document.
  (Right): Acc@1 vs. model size. \edit{Gray curves are per-task scores.}
  }
  \label{fig:query_num}
\end{figure}

\subsection{Analysis of different docid designs}
\label{sec:docid}

Figure~\ref{fig:query_num} compares our proposed unified docid with previous GR docid designs, while keeping all other factors (e.g., query generator, model choice, optimization strategy) constant to ensure an apple-to-apple comparison of docid effectiveness.  
The compared docid designs include:  
(i) \textbf{Random}~\citep{Tay2022TransformerMA}, a baseline that assigns each document a random string as its docid;  
(ii) \textbf{Sentence}~\citep{Bevilacqua2022AutoregressiveSE}, which uses all sentences of each document as its docid;  
(iii) \textbf{Paragraph}~\citep{Tay2022TransformerMA}, which takes the first paragraph of each document as its docid;  
(iv) \textbf{Query}~\citep{Tang2023SemanticEnhancedDS}, which uses a query generator to produce a single query per document as its docid;  
(v) \textbf{Summary}, as introduced in~\citep{li-etal-2024-summarization}, which uses the output of a summarization model as the docid; and 
(vi) \textbf{RQ-VAE}~\citep{Zeng2023ScalableAE}, which trains a RQ-VAE model on document embeddings produced by the \texttt{BGE-Large} model, enabling quantization of document embeddings into a sequence of tokens. This is a widely adopted docid representation in competitive GR systems.

\begin{wrapfigure}[14]{r}{0.6\textwidth}
  \centering
  \vspace{-1em}
  \includegraphics[width=\linewidth]{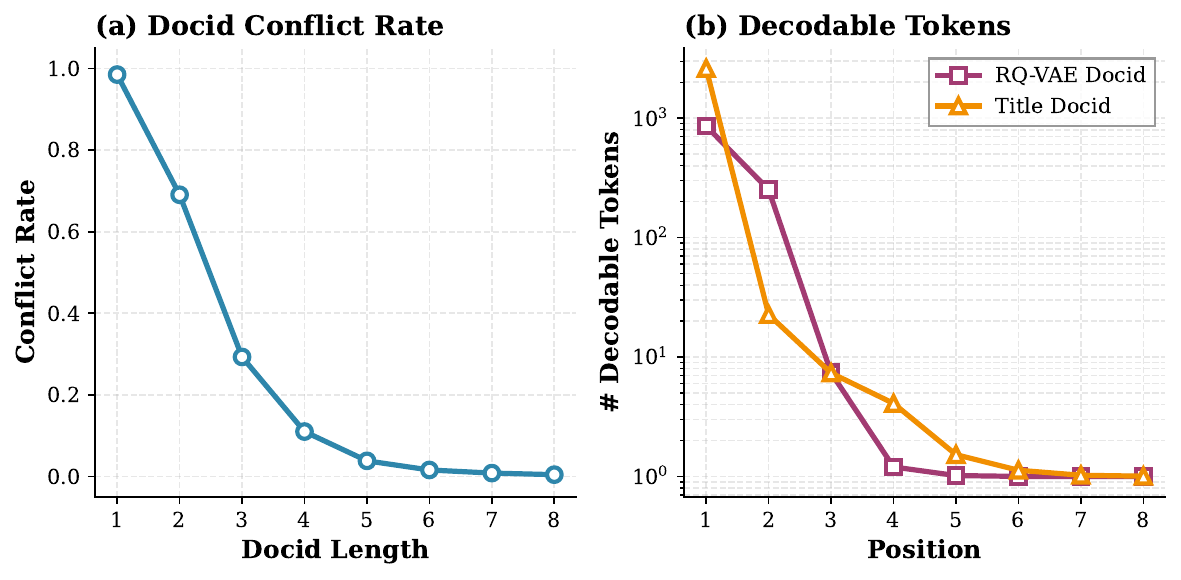}
  \caption{\edit{(Left): Docid conflict rate wrt docid length. (Right): Average number of decodable tokens at each position, for RQ-VAE docid and our title docid.}}\label{docid_analysis}
  \vspace{-2em}
\end{wrapfigure}

\subsection{Analysis of docid distribution}\label{sec:docid-dist}

Figure \ref{docid_analysis} (left) evaluates conflict rates when treating only a prefix of the generated docid as the identifier. The results show that the conflict rate drops below 1\% once the prefix length exceeds six words, and with an eight-word prefix the conflict rate is only 0.45\%. This confirms that our default docid length is sufficient and that retrieval accuracy is not meaningfully impacted by docid length.

Figure \ref{docid_analysis} (right) shows the average number of decodable tokens at each trie step for RQVAE and title-based docids. Title-based docids yield a large branching factor at the first step, with many decodable tokens, followed by a sharp reduction in later steps. In contrast, RQVAE forms a more gradually narrowing trie that becomes nearly deterministic after two to three steps.

\begin{wrapfigure}[14]{r}{0.6\textwidth}
  \centering
  \vspace{-1em}
  \includegraphics[width=\linewidth]{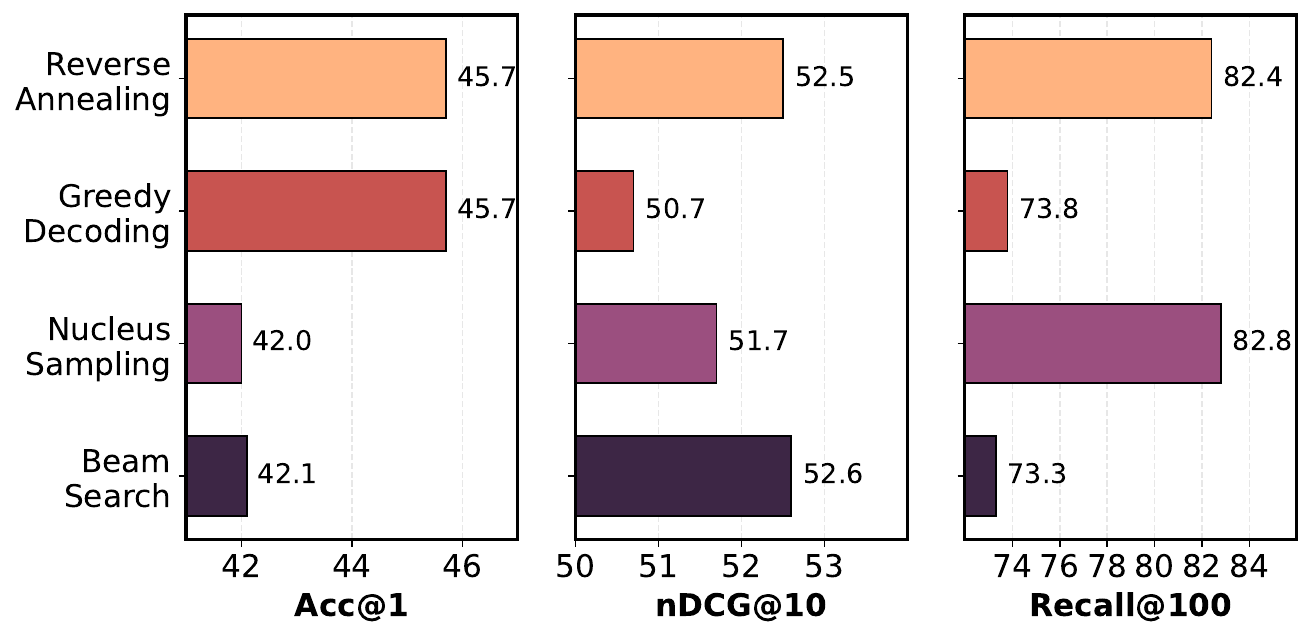}
  \caption{
  Ablation study of decoding algorithms across different metrics. Our proposed reverse annealing decoding achieves a good balance between precision and recall.
  }
  \label{fig:decode}
  \vspace{-2em}
\end{wrapfigure}

\subsection{Analysis of decoding strategies}
\label{sec:decode}

In Figure~\ref{fig:decode}, we compare our reverse annealing decoding with other popular decoding algorithms, including greedy decoding (i.e., greedily sampling from the GR model without replacement), nucleus sampling with a top-p of 0.9, and beam search. All methods decode the top-100 docids for evaluation.
From the results, we observe that greedy decoding achieves the best performance in terms of Acc@1, but lacks diversity and yields low recall. Nucleus sampling performs poorly on Acc@1 but achieves high recall. In contrast, reverse annealing strikes a good balance between precision and recall, achieving competitive results across all metrics.

\subsection{Analysis of query number and model size}
\label{sec:query_num}
The middle section of Figure~\ref{fig:query_num} illustrates the impact of the number of queries generated per document on the average top-1 accuracy of \textsc{ZeroGR}. We observe a clear upward trend: as the number of queries increases, the retrieval performance improves steadily. This highlights the importance of diverse query views for better semantic coverage during indexing. Notably, when using eight queries per document, \textsc{ZeroGR} already reaches performance on par with the strong sparse baseline BM25. Further increasing the query count to sixteen enables \textsc{ZeroGR} to surpass BM25, suggesting that high query diversity provides richer signals for matching user queries to relevant documents.

The right section of Figure~\ref{fig:query_num} examines how the size of the backbone language model affects retrieval performance. For this analysis, we adopt a series of Qwen2.5~\citep{qwen2025qwen25technicalreport} models with varying parameter scales. The results demonstrate a consistent gain in top-1 accuracy on unseen IR tasks as the model size grows, implying that larger models benefit from enhanced generalization and better understanding of the instruction-based retrieval formulation. This finding underscores the value of scaling up model capacity in generative retrieval frameworks, particularly in zero-shot settings.

\begin{table*}[t!]
\centering\small
\setlength\tabcolsep{4pt}
\caption{Combined domain-wise results on MAIR (Acc@1) and BEIR (nDCG@10). Performance of different retrieval models across various domains. See Table~\ref{table:mair} and~\ref{tab:beir} for details.}
\label{tab:main}
\begin{tabular}{l ccccccc ccccc}
\toprule
& \multicolumn{7}{c}{MAIR (38 Tasks)} & \multicolumn{5}{c}{BEIR (11 Tasks)} \\
\cmidrule(lr){2-8} \cmidrule(lr){9-13}
Model &
Avg &
Web. &
Aca. &
Legal &
Med. &
Fin. &
Cod. &
Avg &
Web. &
Aca. &
Med. &
Fin.\\
\midrule
BM25 & \heatblue{26}{36.1} & \heatblue{44}{34.3} & \heatblue{43}{39.2} & \heatblue{29}{34.5} & \heatorange{22}{42.4} & \heatorange{4}{40.0} & \heatblue{100}{17.3} & \heatblue{30}{42.4} & \heatblue{20}{45.4} & \heatblue{37}{38.8} & \heatblue{11}{32.7} & \heatblue{64}{41.6}\\
\midrule
Contriever & \heatblue{39}{33.6} & \heatblue{15}{39.8} & \heatblue{74}{33.4} & \heatblue{69}{26.8} & \heatblue{39}{30.8} & \heatblue{10}{37.3} & \heatblue{98}{17.7} & \heatblue{3}{47.6} & \heatorange{12}{51.5} & \heatblue{15}{43.0} & \heatblue{4}{33.9} & \heatblue{33}{47.6}\\
GTR-T5-base & \heatblue{45}{32.5} & \heatblue{35}{36.0} & \heatblue{73}{33.6} & \heatblue{77}{25.3} & \heatblue{34}{31.9} & \heatblue{9}{37.4} & \heatblue{93}{18.7} & \heatblue{15}{45.3} & \heatorange{8}{50.7} & \heatblue{55}{35.3} & \heatblue{11}{32.7} & \heatblue{45}{45.3}\\
GTR-T5-large & \heatblue{30}{35.4} & \heatblue{15}{39.8} & \heatblue{41}{39.6} & \heatblue{64}{27.8} & \heatblue{34}{31.8} & \heatblue{4}{38.5} & \heatblue{65}{24.0} & \heatblue{1}{48.0} & \heatorange{22}{53.3} & \heatblue{44}{37.4} & \heatblue{7}{33.4} & \heatblue{20}{50.0}\\
\midrule
E5-Base & \heatblue{21}{37.2} & \heatblue{34}{36.2} & \heatorange{6}{48.6} & \heatblue{61}{28.5} & \heatblue{16}{35.3} & \heatorange{30}{44.9} & \heatblue{51}{26.7} & \heatorange{4}{48.9} & \heatorange{14}{51.8} & \heatorange{2}{46.1} & \heatorange{2}{35.0} & \heatblue{19}{50.2}\\
E5-Large & \heatblue{15}{38.2} & \heatblue{22}{38.6} & \heatorange{19}{51.0} & \heatblue{79}{25.0} & \heatblue{14}{35.6} & \heatorange{39}{46.6} & \heatblue{56}{25.7} & \heatorange{6}{49.2} & \heatorange{13}{51.7} & \heatorange{11}{47.9} & \heatorange{14}{37.4} & \heatblue{26}{48.8}\\
BGE-Base & \heatblue{22}{37.0} & \heatblue{22}{38.6} & \heatblue{38}{40.2} & \heatblue{75}{25.8} & \heatblue{4}{37.6} & \heatorange{16}{42.2} & \heatblue{38}{29.0} & \heatorange{13}{50.5} & \heatorange{17}{52.5} & \heatorange{7}{47.1} & \heatorange{7}{36.0} & \heatorange{7}{55.2}\\
BGE-Large & \heatblue{9}{39.4} & \heatblue{17}{39.4} & \heatblue{6}{46.2} & \heatblue{21}{36.0} & \heatblue{6}{37.2} & \heatorange{31}{45.1} & \heatblue{38}{29.0} & \heatorange{19}{51.8} & \heatorange{24}{53.8} & \heatorange{11}{47.9} & \heatorange{18}{38.1} & \heatorange{14}{56.5}\\
\midrule
OpenAI-Embed & \heatblue{3}{40.6} & \heatblue{11}{40.6} & \heatorange{4}{48.2} & \heatblue{47}{31.0} & \heatorange{7}{39.7} & \heatorange{54}{49.4} & \heatblue{40}{28.7} & \heatorange{32}{54.2} & \heatorange{37}{56.3} & \heatorange{7}{47.2} & \heatorange{15}{37.6} & \heatorange{51}{63.4}\\
E5-mistral-7B & \heatorange{30}{46.8} & \heatorange{14}{45.4} & \heatorange{42}{55.4} & \heatorange{12}{42.3} & \heatorange{25}{43.1} & \heatorange{85}{55.3} & \heatorange{19}{40.0} & \heatorange{40}{55.7} & \heatorange{38}{56.4} & \heatorange{15}{48.6} & \heatorange{26}{39.6} & \heatorange{79}{68.8}\\
GritLM-7B & \heatorange{31}{47.0} & \heatorange{7}{44.1} & \heatorange{57}{58.2} & \heatorange{17}{43.3} & \heatorange{23}{42.6} & \heatorange{97}{57.6} & \heatorange{19}{40.0} & \heatblue{16}{45.0} & \heatblue{8}{47.7} & \heatorange{13}{48.2} & \heatorange{12}{36.9} & \heatblue{84}{37.8}\\
\midrule
ZeroGR-3B & 41.1 & 42.7 & 47.4 & 40.0 & 38.3 & 39.2 & 36.3 & 48.1 & 49.2 & 45.8 & 34.7 & 53.8\\
\bottomrule
\end{tabular}
\end{table*}

\begin{table}
\centering\small
\setlength\tabcolsep{5pt}
\captionof{table}{Performance of different generative retrieval models across various datasets on BEIR.}
    \label{tab:beir_small}
      \begin{tabular}{l l ccccccc}
      \toprule
      Method & Training data & Avg & Argu. & SciF. & NFC. & FiQA & SciD. & Covid \\
      \midrule
            GENRE~\citep{DeCao2020AutoregressiveER}  & GPL 
        & \heatblue{37}{23.0} 
        & \heatorange{12}{\textbf{42.5}} 
        & \heatblue{52}{42.3} 
        & \heatblue{25}{20.0} 
        & \heatblue{38}{11.6} 
        & \heatblue{20}{6.8} 
        & \heatblue{100}{14.7} \\
      GENRET~\citep{Sun2023LearningTT} & GPL 
        & \heatblue{6}{41.1} 
        & \heatblue{2}{34.3} 
        & \heatblue{15}{63.9} 
        & \heatblue{5}{31.6} 
        & \heatblue{7}{30.2} 
        & \heatblue{6}{14.9} 
        & \heatblue{3}{71.8} \\
      GLEN~\citep{Lee2023GLENGR} & NQ320k  
        & --   
        & \heatblue{30}{17.6} 
        & --   
        & \heatblue{32}{15.9} 
        & --   
        & --   
        & --   \\
      TIGER~\citep{Rajput2023RecommenderSW}  &  OpenInstIR 
        &  \heatblue{24}{31.0} 
        & \heatblue{36}{14.0} 
        & \heatblue{61}{37.0} 
        & \heatorange{8}{\textbf{39.5}} 
        & \heatblue{31}{16.0} 
        & \heatblue{8}{14.0} 
        & \heatblue{13}{65.7} \\
        \midrule
      ZeroGR (Ours) & OpenInstIR 
        & \textbf{44.9} 
        & 35.4
        & \textbf{72.8} 
        & 34.7 
        & \textbf{34.1} 
        & \textbf{18.7} 
        & \textbf{73.5} \\
      \bottomrule
      \end{tabular}
    \vspace{-0.1cm}
\end{table}

\section{Benchmark Evaluation}

\subsection{Evaluation results on MAIR}
\label{sec:mair}

As shown in Table~\ref{tab:main} (MAIR), our proposed \textsc{ZeroGR} framework demonstrates strong performance across a wide range of retrieval tasks. It achieves an average score of 41.1 (Acc@1), substantially outperforming traditional sparse retrieval methods like BM25 and widely adopted dense retrieval models such as Contriever, GTR, E5, BGE, and on par with instruction-tuned OpenAI-Embedding-v3-Small. These results highlight the effectiveness of our instruction-based generative retrieval approach in capturing deeper semantic relevance. Our detailed experimental results on MAIR are shown in Table~\ref{table:mair} (Appendix).

The performance gains of \textsc{ZeroGR} are not limited to familiar tasks but also generalize well to unseen domains. \edit{Notably, the model performs better than baselines on several previously unseen datasets, including Apple, MB, PM.A, DD, and NCL (see Table~\ref{table:mair}, Appendix).} This demonstrates the robustness and transferability of the approach, as it adapts effectively to new retrieval settings without requiring additional task-specific supervised data.

\edit{Using a 3B LLM,
\textsc{ZeroGR} can achieve strong performance across different tasks compared to baselines, though it still underperforms large embedding models such as GritLM-7B and E5-Mistral-7B.}
This indicates that our design is highly parameter-efficient, achieving strong performance across diverse tasks without relying on massive model scaling. 

\subsection{Evaluation results on BEIR}
\label{sec:beir}

\edit{As shown in Table~\ref{tab:main} (BEIR), \textsc{ZeroGR} outperforms several baselines such as BM25, Contriever, GTR, and GritLM-7B, but still underperforms other dense retrieval methods.}
We have more detailed comparison between ZeroGR and dense retrieval on BEIR in Appendix~\ref{sec:all_results}.
Furthermore, Table~\ref{tab:beir_small} compares \textsc{ZeroGR} with previous generative retrieval baselines on BEIR, which we can see our method achieves best performance among most datasets.

\section{Conclusion}

We have presented \textsc{ZeroGR}, an instruction-driven framework that extends generative retrieval to zero-shot scenarios. 
By unifying three key components, viz.\ a model-based docid generator, an instruction-conditioned query generator, and a reverse-annealed decoding algorithm, \textsc{ZeroGR} transforms a corpus and a natural-language task description into a task-specific generative index without requiring supervision. 
Systematic ablation studies along task diversity, query volume, and model size reveal consistent performance improvements. 
Empirical evaluations on MAIR tasks and BEIR datasets demonstrate the effectiveness of \textsc{ZeroGR}.

\edit{The limitations of this work include a lack of evaluation on large-scale corpora (e.g., those with over 1M documents) and the use of relatively small LLMs (our largest model is only 3B). We believe future work is required to scale both the corpus size and the model size.}

\subsubsection*{Acknowledgements}
We thank our reviewers for their helpful feedback. This research was (partially) supported by the Dutch Research Council (NWO), under project numbers 024.004.022, NWA.1389.20.\-183, and KICH3.LTP.20.006, and the European Union under grant agreements No.\ 101070212 (FINDHR) and No.\ 101201510 (UNITE).
Views and opinions expressed are those of the author(s) only and do not necessarily reflect those of their respective employers, funders and/or granting authorities.

\bibliographystyle{plainnat}
\bibliography{ref,mair_ref}

\appendix

\newpage

\section{OpenInstIR}\label{sec:train-data}
We collect the training set OpenInstIR based on the training splits of published information retrieval datasets (e.g., these from \citep{Welleck2021NaturalProofsMT,Chen2022TargetawareAR,Hendrycks2021MeasuringMP,Chen2022TargetawareAR,Chen2022TargetawareAR,Cohan2020SPECTERDR,Muennighoff2023OctoPackIT,Husain2019CodeSearchNetCE,Austin2021ProgramSW,Hendrycks2021MeasuringCC,Liu2023RepoBenchBR,Zhu2022TowardsCD,Maia2018WWW18OC,Chalkidis2021RegulatoryCT,Chen2022ConvFinQAET,Li2023LeCaRDv2AL,Kornilova2019BillSumAC,Hendrycks2021CUADAE,Boteva2016AFL,Levy2017ZeroShotRE,Hoffart2011RobustDO,Adlakha2021TopiOCQAOC,Zhang2024ExcluIREN,Dinan2018WizardOW}).
Specifically, we pair each query with its corresponding positive document and annotate the pairs with natural language instructions following MAIR, forming standardized (instruction, query, document) triples entirely from human annotations.
Table~\ref{tab:domain_stats} reports statistics of OpenInstIR, which spans 6 domains and 66 tasks; during training, we upsample tasks with fewer than 5K samples and downsample those with more than 500K samples.
Table~\ref{tab:doc_type} compares query and document types between OpenInstIR and the evaluation set, showing 9 shared query types and 14 shared document types, while the test set contains 9 additional query types and 8 additional document types unseen during training.

\begin{table}[h]
\centering\footnotesize
\caption{\edit{OpenInstIR dataset statistics by domain.}}
\label{tab:domain_stats}
\setlength{\tabcolsep}{5pt}
\begin{tabular}{l r l r l r l r}
\toprule
Dataset & Size & Dataset & Size & Dataset & Size & \textbf{Dataset} & Size \\
\midrule
\multicolumn{8}{c}{\emph{Academic}} \\
\midrule
S2-TC & 100K & S2-AC & 100K & S2-TA & 100K & TAD & 208K \\
TAS2 & 108K & StackMath & 47K & Proof-P & 16K & Proof-R & 2K \\
Stacks-P & 11K & Stacks-R & 9K & CompMath & 8K & SciDocs & 1K \\
SciFact & 809 & LitSearch & 146 &  &  &  &  \\
\midrule
\multicolumn{8}{c}{\emph{Code}} \\
\midrule
CSN & 1.88M & CodeEdit & 21K & SWE-B & 19K & RepoBench & 17K \\
HF-API & 8K & TLDR & 6K & TensorAPI & 6K & APPS & 5K \\
LeetCode & 2K & CoNaLa & 2K & TorchAPI & 837 & HumanEval-X & 720 \\
MBPP & 374 &  &  &  &  &  &  \\
\midrule
\multicolumn{8}{c}{\emph{Finance}} \\
\midrule
USNews & 10K & FinQA & 6K & FiQA & 6K & HC3-Fin & 3K \\
ConvFinQA & 3K & Goldman & 2K & TAT-DQA & 1K & TradeEvt & 900 \\
\midrule
\multicolumn{8}{c}{\emph{Legal}} \\
\midrule
LePaRD & 22.7M & CLERC & 327K & BillSum & 19K & REGIR-UK2EU & 2K \\
REGIR-EU2UK & 2K & BSARD & 886 & CUAD & 717 &  &  \\
\midrule
\multicolumn{8}{c}{\emph{Medical}} \\
\midrule
PMQA-C & 197K & PMQA-A & 197K & Huatuo & 25K & NFCorpus & 3K \\
CARE & 77 &  &  &  &  &  &  \\
\midrule
\multicolumn{8}{c}{\emph{Web}} \\
\midrule
Reddit & 12.7M & AGNews & 1.16M & CC-News & 708K & XSum & 204K \\
zsRE & 148K & Fever & 110K & ToT & 109K & WoW & 64K \\
TopiOCQA & 45K & AY2 & 18K & CQADup & 13K & InstructIR & 10K \\
Quora & 10K & WnCw & 5K & TREx & 5K & ExcluIR & 3K \\
NevIR & 2K & ArguAna & 1K &  &  &  &  \\
\bottomrule
\end{tabular}
\end{table}

\begin{table}[htbp]
\centering\footnotesize
\caption{Query and doc types between our train and eval set.}\label{tab:doc_type}
\begin{tabular}{@{}p{7cm}p{7cm}@{}}
\toprule
Query types & Document types \\

\midrule
\multicolumn{2}{c}{\emph{Training $\cap$ Evaluation}} \\
\midrule
Question, Dialog, Claim, Function Header, NL Command, Code Problem, Math question, Paper Title, Summary
& Document, Answer, Function, Command Doc, Solution, Article, Articles, Medical Doc, Paragraph, Page, Statute, Table\\
(\textit{9 types}) & (\textit{14 types}) \\

\midrule
\multicolumn{2}{c}{\emph{Only in Evaluation}} \\
\midrule
Health Record, Topic, Situation, Request, Patient Data \& Description, Medical Case, Medical Claim, Numerical Claim
& Clinical Trials, Prior Case, Communications, Dataset, Music, Tweet, News, POI, Table\\
(\textit{9 types}) & (\textit{8 types}) 
\\
\midrule
\multicolumn{2}{c}{\emph{Only in Training}} \\
\midrule
Math Statement, Entity \& Relation, Paper Abstract, Entity Mention, CNL Command, GitHub Issue, Commit, Code Context, Math Question, Title, EU Directive, UK Legislation, Instruction, Reaction, Description 
& Entity Page, Citation, Proof, Reference, Duplicate Question, Related File, Code Diff, Next Function, HuggingFace API, Tensor API, PyTorch API,  UK Legislation, EU Directive, Highlight, Proteins Documents, Wikipedia Page
\\
(\emph{14 types}) & (\emph{16 types})\\
\bottomrule
\end{tabular}
\end{table}

\begin{table}[t]
\centering\footnotesize
\caption{\edit{Development set for ablation study.}}\label{tab:dev-set}
\begin{tabular}{p{0.2\textwidth}p{0.2\textwidth}p{0.5\textwidth}}
\toprule
Name & Model & Task List (with query count) \\
\midrule
\raggedright Figure~2, Figure~3 (left) & 
Llama-1B & 
\{ToolBench (100), AILA2019-Case (50), NFCorpus (100), SciFact (100), ArguAna (100), LitSearch (100), ClinicalTrials\_2023 (37), FinanceBench (100), SciDocs (100), News21 (100), TopiOCQA (100), Touche (49), FiQA (100)\}\\
\midrule
\raggedright Figure~3 (middle, right) &
\raggedright Llama-1B, or Qwen2.5 & 
\{LeetCode (100), Competition-Math (100), TMDB (100), Stein\_Proof (64), PytorchAPI (100)\} \\
\midrule
Figure~\ref{fig:decode} & 
\raggedright Llama-1B & 
\{Leetcode (100), Math (100), BillSum (100), SciFact (100), TAT-DQA (70), ConvFinQA (96)\}\\
\bottomrule
\end{tabular}
\end{table}

\begin{table}[t]
\centering\footnotesize
\caption{\edit{Examples of different types of docids.}}
\label{tab:docid_examples}
\begin{tabular}{l p{10cm}}
\toprule
\textbf{Type} & \textbf{Example} \\
\midrule
Random & asd8xc2c9ma90xj2398 \\
\midrule
Sentence & LIMASSOL, Cyprus, April 28, 2021 /PRNewswire/ -- One of the top financial investment firms of the FX industry, Windsor Brokers .... \\
\midrule
Paragraph & LIMASSOL, Cyprus, April 28, 2021 /PRNewswire/ -- One of the top financial investment firms of the FX industry, Windsor Brokers .... \\
\midrule
Query & Induction of myelodysplasia by myeloid-derived suppressor cells. \\
\midrule
Summary & 1. Game of Thrones season 7 2. Plot and storyline 3. New cast members 4. Filming locations 5. Critical reception and ratings \\
\midrule
RQ-VAE & \texttt{<|g16289|> <|g13509|> <|g10485|> <|g11274|> <|g369|> <|g3661|> <|g13026|> <|g8187|>} \\
\midrule 
IDF & brokerswindsor mt4 brokerswere kontos windsorbrokers \\ 
\midrule
Ours & rna folding computational methods thermodynamic optimization\\
\bottomrule
\end{tabular}
\end{table}

\section{Ablation Study Details}
To reduce computational overhead, our ablation studies are conducted on the Dev subset rather than the full test set; Table~\ref{tab:dev-set} provides the task breakdown.
For the docid design ablation (Section~\ref{sec:docid}), Table~\ref{tab:docid_examples} presents representative docids, from which we observe that sentence, paragraph, and IDF docids are often noisy; query and summary docids fail to capture key concepts early and are suboptimal for left-to-right GR decoding; and RQ-VAE does not exploit the language generation capabilities of LLMs.
For the decoding ablation study, we fix the beam size to 100 and use the same hyperparameters as in the main experiments. No top-k or top-p pruning is applied, and the reverse annealing parameter is kept consistent across all datasets and models.

\begin{table}[t]
    \centering
    \setlength{\tabcolsep}{0.5pt}
    \tiny
    \caption{Model performance (top-1 retrieval accuracy) on seen and unseen subset of MAIR.}
    \label{table:mair}
    \begin{tabular}{lrrrrrrrrrrrrrrrrrrrr}
    \toprule
    Dataset &  & \multicolumn{15}{c}{Seen Subset} & \multicolumn{4}{c}{Unseen Subset} \\
    \cmidrule{3-17} 
    \cmidrule(l){18-21}
    Model & Avg & FiQA & NFC.s & SciD. & SciF. & ToQA. & TAT & CoF. & LeetC. & LitSe. & BiSum & CodeSe. & Math & ConvF. & Conala & StMath & Apple & FinBen & AILAC & AILAS \\
    \midrule
    BM25 & 36.1 & 24.0 & 45.5 & 16.0 & 53.0 & 11.0 & 67.1 & 47.9 & 12.0 & 66.0 & 69.0 & 33.0 & 41.0 & 47.9 & 7.0 & 20.0 & 52.1 & 9.0 & 14.0 & 10.0 \\
    \midrule
    Contriever & 33.6 & 33.0 & 43.5 & 17.0 & 62.0 & 11.0 & 54.3 & 42.7 & 7.0 & 39.0 & 54.0 & 37.0 & 36.0 & 42.7 & 9.0 & 13.0 & 50.7 & 6.0 & 12.0 & 6.0 \\
    GTR-T5-base & 32.5 & 33.0 & 43.0 & 12.0 & 50.0 & 16.0 & 58.6 & 37.5 & 6.0 & 41.0 & 47.0 & 41.0 & 49.0 & 37.5 & 9.0 & 16.0 & 47.9 & 10.0 & 4.0 & 10.0 \\
    GTR-T5-large & 35.4 & 45.0 & 43.5 & 14.0 & 55.0 & 12.0 & 40.0 & 42.7 & 10.0 & 43.0 & 59.0 & 51.0 & 63.0 & 42.7 & 11.0 & 23.0 & 50.7 & 14.0 & 6.0 & 6.0 \\
    E5-Base & 37.2 & 41.0 & 40.0 & 17.0 & 63.0 & 16.0 & 61.4 & 52.1 & 11.0 & 49.0 & 61.0 & 59.0 & 78.0 & 52.1 & 10.0 & 36.0 & 52.1 & 18.0 & 6.0 & 12.0 \\
    E5-Large & 38.2 & 45.0 & 45.5 & 20.0 & 67.0 & 13.0 & 70.0 & 57.3 & 9.0 & 49.0 & 52.0 & 57.0 & 75.0 & 57.3 & 11.0 & 44.0 & 47.9 & 13.0 & 12.0 & 6.0 \\
    BGE-Base & 37.0 & 43.0 & 43.5 & 20.0 & 62.0 & 13.0 & 58.6 & 41.7 & 10.0 & 44.0 & 67.0 & 64.0 & 50.0 & 41.7 & 13.0 & 25.0 & 47.9 & 20.0 & 8.0 & 8.0 \\
    BGE-Large & 39.4 & 51.0 & 46.5 & 22.0 & 65.0 & 13.0 & 60.0 & 44.8 & 13.0 & 56.0 & 68.0 & 66.0 & 66.0 & 44.8 & 8.0 & 22.0 & 46.6 & 23.0 & 8.0 & 8.0 \\
    \midrule
    OpenAI-Embed & 40.6 & 51.0 & 51.0 & 22.0 & 60.0 & 19.0 & 62.9 & 51.0 & 6.0 & 53.0 & 59.0 & 67.0 & 73.0 & 51.0 & 13.0 & 33.0 & 52.1 & 30.0 & 10.0 & 10.0 \\
    GTE-Qwen2-1.5B & 44.4 & 54.0 & 50.0 & 24.0 & 69.0 & 25.0 & 65.7 & 65.6 & 41.0 & 63.0 & 79.0 & 70.0 & 84.0 & 65.6 & 20.0 & 40.0 & 47.9 & 33.0 & 12.0 & 10.0 \\
    E5-mistral-7B & 46.8 & 60.0 & 50.5 & 17.0 & 67.0 & 14.0 & 67.1 & 64.6 & 36.0 & 68.0 & 74.0 & 54.0 & 78.0 & 64.6 & 30.0 & 47.0 & 43.8 & 41.0 & 12.0 & 38.0 \\
    GritLM-7B & 47.0 & 63.0 & 49.5 & 29.0 & 69.0 & 17.0 & 85.7 & 62.5 & 46.0 & 60.0 & 74.0 & 53.0 & 87.0 & 62.5 & 21.0 & 46.0 & 43.8 & 33.0 & 12.0 & 42.0 \\
    \midrule
    \textbf{ZeroGR-3B} & 41.1 & 37.0 & 36.5 & 24.0 & 51.0 & 13.0 & 38.6 & 57.3 & 36.0 & 41.0 & 81.0 & 61.0 & 81.0 & 57.3 & 12.0 & 40.0 & 52.1 & 11.0 & 12.0 & 22.0 \\    
    \toprule
    & \multicolumn{20}{c}{\emph{Unseen subset}} \\
    \cmidrule{2-21}
    & ACOR. & CPCD & CORE & MB. & PM. & PM.A & CliDS & CliT23 & DD & Table & QuanT & PoRec & Monant & NCL. & NCL.T & Legal & Geno. & Touche & CliT21 & News21 \\
    \midrule
    BM25 & 32.8 & 1.0 & 37.5 & 83.8 & 53.9 & 6.5 & 28.3 & 51.4 & 15.6 & 10.0 & 86.9 & 24.5 & 67.4 & 50.7 & 22.2 & 45.0 & 52.8 & 59.2 & 33.3 & 10.9 \\
    \midrule
    Contriever & 40.4 & 1.0 & 52.5 & 89.2 & 32.9 & 0.0 & 6.7 & 37.8 & 21.3 & 8.3 & 76.8 & 46.7 & 65.0 & 60.0 & 34.2 & 35.0 & 27.8 & 52.0 & 32.7 & 23.8 \\
    GTR-T5-base & 31.3 & 1.0 & 47.5 & 89.2 & 36.8 & 1.6 & 13.3 & 36.5 & 13.6 & 12.5 & 77.8 & 37.7 & 70.0 & 48.0 & 22.2 & 40.0 & 25.0 & 55.1 & 29.3 & 15.6 \\
    GTR-T5-large & 34.8 & 3.0 & 60.0 & 91.9 & 31.6 & 1.6 & 8.3 & 39.2 & 16.2 & 10.0 & 78.8 & 48.7 & 68.0 & 53.3 & 23.9 & 40.0 & 25.0 & 65.3 & 37.3 & 19.1 \\
    E5-Base & 40.4 & 3.0 & 42.5 & 81.1 & 43.4 & 6.5 & 11.7 & 39.2 & 13.2 & 5.8 & 78.8 & 54.2 & 72.0 & 55.3 & 27.4 & 35.0 & 33.3 & 37.8 & 36.0 & 14.5 \\
    E5-Large & 38.4 & 3.0 & 45.0 & 86.5 & 36.8 & 4.8 & 15.0 & 40.5 & 12.3 & 7.5 & 80.8 & 52.5 & 71.0 & 55.3 & 47.9 & 30.0 & 38.9 & 41.8 & 32.0 & 17.2 \\
    BGE-Base & 39.4 & 0.0 & 45.0 & 91.9 & 48.7 & 0.0 & 30.0 & 31.1 & 16.4 & 8.3 & 81.8 & 44.4 & 70.0 & 57.3 & 42.7 & 20.0 & 36.1 & 41.8 & 41.3 & 19.9 \\
    BGE-Large & 36.9 & 0.0 & 52.5 & 94.6 & 42.1 & 3.2 & 23.3 & 28.4 & 17.8 & 5.0 & 81.8 & 50.4 & 74.0 & 54.0 & 40.2 & 60.0 & 38.9 & 51.0 & 41.3 & 15.2 \\
    \midrule
    OpenAI-Embed & 32.8 & 1.0 & 55.0 & 86.5 & 46.1 & 3.2 & 25.0 & 44.6 & 13.5 & 12.5 & 86.9 & 47.3 & 76.0 & 57.3 & 49.6 & 45.0 & 33.3 & 52.0 & 38.0 & 13.7 \\
    GTE-Qwen2-1.5B & 37.9 & 5.1 & 70.0 & 81.1 & 14.5 & 8.1 & 20.0 & 17.6 & 14.8 & 14.2 & 85.9 & 58.6 & 74.2 & 61.3 & 51.3 & 40.0 & 61.1 & 65.3 & 31.3 & 23.0 \\
    E5-mistral-7B & 41.9 & 5.0 & 60.0 & 83.8 & 43.4 & 1.6 & 46.7 & 48.6 & 19.4 & 11.7 & 83.8 & 66.1 & 71.0 & 62.7 & 58.1 & 45.0 & 36.1 & 58.2 & 46.7 & 25.4 \\
    GritLM-7B & 35.4 & 7.0 & 65.0 & 70.3 & 59.2 & 0.0 & 28.3 & 45.9 & 14.8 & 10.8 & 86.9 & 72.2 & 77.0 & 63.3 & 50.4 & 45.0 & 33.3 & 57.1 & 47.3 & 22.7 \\
    \midrule
    \textbf{ZeroGR-3B} & 26.7 & 4.0 & 55.0 & 89.2 & 23.9 & 12.9 & 25.0 & 37.9 & 44.4 & 12.0 & 79.8 & 60.9 & 69.7 & 65.3 & 36.2 & 45.0 & 58.3 & 46.9 & 42.0 & 21.5 \\
    
    \bottomrule
    \end{tabular}
    \end{table}
\begin{table*}[t]
\centering\small
\setlength{\tabcolsep}{2.5pt}
\caption{nDCG@10 on BEIR benchmark datasets.}
\label{tab:beir}
\begin{tabular}{llcccccc}
\toprule
Category & Method & Avg. & ArguAna & SciFact & NFCorpus & FiQA & SciDocs \\
\midrule
Sparse & BM25 & 42.3 & 32.7 & 65.1 & 32.7 & 24.8 & 12.4 \\
\midrule
DR & Contriever & 47.6 & 32.1 & 70.3 & 33.9 & 35.5 & 15.7 \\
DR & GTR-T5-base & 45.3 & 32.7 & 58.6 & 32.7 & 34.5 & 12.1 \\
DR & GTR-T5-large & 48.0 & 34.3 & 61.9 & 33.4 & 43.3 & 12.8 \\
DR & E5-Base & 48.9 & 31.1 & 73.9 & 35.0 & 39.6 & 18.3 \\
DR & E5-Large & 49.2 & 31.7 & 76.3 & 37.4 & 42.3 & 19.6 \\
DR & BGE-Base & 50.5 & 41.8 & 74.3 & 36.0 & 43.4 & 19.8 \\
DR & BGE-Large & 51.8 & 41.6 & 75.2 & 38.1 & 48.5 & 20.6 \\
DR & E5-mistral-7B & 55.7 & 44.1 & 76.6 & 39.6 & 59.7 & 20.7 \\
DR & GritLM-7B & 45.0 & 40.7 & 76.8 & 36.9 & 44.1 & 19.6 \\
DR & OpenAI Embed & 54.2 & 37.1 & 73.1 & 37.6 & 48.5 & 21.2 \\
\midrule
GR & GENRE & -- & 42.5 & 42.3 & 20.0 & 11.6 & 6.8 \\
GR & GENRET & -- & 34.3 & 63.9 & 31.6 & 30.2 & 14.9 \\
GR & GLEN & -- & 17.6 & -- & 15.9 & -- & -- \\
GR & TIGER (Llama-3B) & -- & 14.0 & 37.0 & 39.5 & 16.0 & 14.0 \\
GR & ZeroGR-3B & 48.1 & 35.4 & 72.8 & 34.7 & 34.1 & 18.7 \\
\midrule
Category & Method & Touche & TREC-News & Fever & Quora & Covid & CQADupStack \\
\midrule
Sparse & BM25 & 59.0 & 20.7 & 58.3 & 73.8 & 58.3 & 28.0 \\
\midrule
DR & Contriever  & 42.5 & 27.3 & 90.6 & 86.6 & 59.6 & 29.9 \\
DR & GTR-T5-base & 48.1 & 22.5 & 83.2 & 88.7 & 56.1 & 28.9 \\
DR & GTR-T5-large & 53.1 & 26.6 & 86.8 & 89.1 & 56.7 & 29.8 \\
DR & E5-Base & 41.1 & 22.9 & 91.1 & 86.4 & 60.7 & 38.3 \\
DR & E5-Large & 34.8 & 25.3 & 93.1 & 86.9 & 55.2 & 38.5 \\
DR & BGE-Base & 41.4 & 21.2 & 85.6 & 89.8 & 67.1 & 35.1 \\
DR & BGE-Large & 45.5 & 21.4 & 86.6 & 89.3 & 64.5 & 38.3 \\
DR & E5-mistral-7B & 46.8 & 29.4 & 91.8 & 84.8 & 77.8 & 41.4 \\
DR & GritLM-7B & 21.5 & 34.9 & 68.9 & 84.9 & 31.5 & 35.0 \\
DR & OpenAI Embed & 47.5 & 26.2 & 92.8 & 89.9 & 78.2 & 44.1 \\
\midrule
GR & GENRE & -- & -- & -- & -- & 14.7 & -- \\
GR & GENRET & -- & -- & -- & -- & 71.8 & -- \\
GR & GLEN & -- & -- & -- & -- & -- & -- \\
GR & TIGER (Llama-3B) & 58.1 & 16.4 & -- & 59.6 & 65.7 & -- \\
GR & ZeroGR-3B & 37.5 & 23.5 & 86.7 & 76.7 & 73.5 & 35.2 \\
\bottomrule
\end{tabular}
\end{table*}

\section{Benchmark Results}\label{sec:all_results}

On BEIR, our method underperforms state-of-the-art dense retrieval models (Tables~\ref{table:mair} and~\ref{tab:beir}). We attribute this gap to two main factors. First, many BEIR tasks overlap with the large-scale training data used by modern embedding models, whereas MAIR is a newer and more diverse benchmark on which these models generalize less effectively, giving dense retrieval an inherent advantage on BEIR. Second, dense retrieval has benefited from years of targeted optimization for zero-shot evaluation, including hard-negative mining, distillation, and large-scale pre-training~\cite{Xiong2020ApproximateNN,Wang2022TextEB}. In contrast, our work represents one of the first efforts to systematically improve zero-shot generative retrieval across heterogeneous tasks using a simple training objective (Eq.~\ref{eq:ce-loss}), making a performance gap with mature dense systems expected. Nevertheless, our approach substantially narrows this gap compared to prior generative retrieval methods. Notably, while our query generator is trained on diverse data, the retrieval model itself is initialized from Llama-3B-Instruct and trained solely on synthetic data, without supervision from IR datasets; incorporating IR-specific pre-training or advanced techniques such as hard-negative mining and distillation may further reduce the remaining gap.

\begin{table}[t]
\centering\footnotesize
\setlength\tabcolsep{4pt}
\caption{Estimated indexing and retrieval cost.}
\label{tab:efficiency_comparison}
\begin{tabular}{l cl cc}
\toprule
Method & \multicolumn{2}{c}{Indexing (offline)} & \multicolumn{2}{c}{Retrieval (online per query)} \\
\cmidrule(r){2-3}
\cmidrule{4-5}
 & FLOPs & Est.\ time & FLOPs & Latency \\ 
 \midrule
Dense-3B~\citep{Wang2023ImprovingTE} & $6.00 \times 10^{17}$ & ${\sim} 8.0$ min & $3.48 \times 10^{12}$ & ${\sim} 2.8$ ms \\
GENRET-3B~\citep{Sun2023LearningTT} & $4.60 \times 10^{19}$ & ${\sim} 10.2$ h & $1.34 \times 10^{12}$ & ${\sim} 1.1$ ms \\
Summary-3B~\citep{li-etal-2024-summarization} & $1.15 \times 10^{19}$ & ${\sim} 2.6$ h & $4.80 \times 10^{12}$ & ${\sim} 3.8$ ms \\
ZeroGR-3B & $8.06 \times 10^{18}$ & ${\sim} 1.8$ h & $1.34 \times 10^{12}$ & ${\sim} 1.1$ ms \\ 
\bottomrule
\end{tabular}%
\end{table}

\section{Model Efficiency}
Table~\ref{tab:efficiency_comparison} reports the estimated computational cost of dense retrieval and generative retrieval methods~\citep{Wang2023ImprovingTE,Rajput2023RecommenderSW,li-etal-2024-summarization} under a fixed setting (Llama-3B, single A100 GPU with vLLM) on a 100K-document corpus for top-($K$=1) retrieval (noting that GR slows with larger $K$, while Dense slows with larger corpora).
ZeroGR is more efficient than prior GR methods due to its compact docid design and reduced training epochs.
Compared with dense retrieval, GR incurs a much higher one-time indexing cost due to model training, which can be amortized during inference.

\section{Prompts}\label{sec:id_prompt}

{\footnotesize
\begin{verbatim}
1. **Length**: Strictly 6-8 words (terms/words)  
2. **Term Inclusion**: Must include 3-5 core terms directly from 
the document  
3. **Term Positioning**: Rank by relevance and importance (highest 
→ lowest, general → specific)
4. **Formatting**:  
   - Use lowercase letters, numbers, and spaces only  
   - Preserve special terms/symbols (e.g., PD3.1)  
   - **No articles** (a, the), **linking verbs**, or auxiliary 
verbs  
   - **No verbs** (use nouns/adjectives only)  
5. **Requirements**:  
   - Terms must be derivable from the document
   - Ensure uniqueness and precise core content representation  
\end{verbatim}
}

\section{Additional Results}
Table \ref{tab:baseline-info} lists the size, training data, and links of the baselines. Table \ref{tab:gr-info} shows the size, training data, docid type, and decoding strategy of the generative retrieval models.

We conduct an ablation study in which we replace the doc2query-generated queries in the baseline TIGER (RQ-VAE) model with queries from our generator. As reported in Table \ref{tab:ablate-q2d}, using our queries alone yields an overall improvement of +6.4. The gains are especially pronounced on domain-specific benchmarks, including legal (BillSum +30.0), finance (FinQA +26.7, TAT-DQA +20.0), code (LeetCode +24.0, CodeSearchNet +23.0), and scientific retrieval (Competition-Math +21.0, LitSearch +19.0). These results indicate that high-quality, diverse queries are crucial, particularly for specialized retrieval tasks.

\section{Use of Large Language Models}

In the preparation of this manuscript, we used large language models (LLMs) to enhance the clarity and linguistic quality of our academic writing. Specifically, we employed Claude Sonnet 4 (Anthropic) and GPT-5 (OpenAI) for language polishing and refinement purposes.

\begin{table}[t]
\centering\footnotesize
\setlength{\tabcolsep}{2pt}
\caption{\edit{Dense retrieval model information.}}\label{tab:baseline-info}
\begin{tabular}{lcll}
\toprule
Model & Size & Training Data & Link \\
\midrule
BM25 & N/A & N/A & https://github.com/cvangysel/BM25S \\
Contriever-MARCO & 110M & MS MARCO & https://github.com/facebookresearch/contriever \\
GTR-base & 110M & MS MARCO & https://huggingface.co/google/gtr-base \\
GTR-large & 335M & MS MARCO & https://huggingface.co/google/gtr-large \\
E5-base & 110M & unknown & https://huggingface.co/intfloat/e5-base-v2 \\
E5-large & 335M & unknown & https://huggingface.co/intfloat/e5-large-v2 \\
BGE-base & 110M & MTEB-Train & https://huggingface.co/BAAI/bge-base-en-v1.5 \\
BGE-large & 335M & MTEB-Train & https://huggingface.co/BAAI/bge-large-en-v1.5 \\
OpenAI-Embed-Small & unknown & unknown & https://platform.openai.com/docs/guides/embeddings \\
E5-Mistral-7B-instruct & 7B & E5 (LLM generated) & https://huggingface.co/intfloat/e5-mistral-7b-instruct \\
GritLM-7B & 7B & E5 (LLM generated) & https://huggingface.co/GritLM/GritLM-7B \\
\bottomrule
\end{tabular}
\end{table}

\begin{table}
\centering\small
\setlength\tabcolsep{5.5pt}
\captionof{table}{\edit{Generative retrieval model information.}}\label{tab:gr-info}
\begin{tabular}{lllll}
\toprule
Method & Training Data & Model Size & DocID Type & Decoding \\
\midrule
GENRE~\citep{DeCao2020AutoregressiveER}  & GPL & T5-220M & Title & Beam Search\\
GENRET~\citep{Sun2023LearningTT} & GPL & T5-220M & RQ-VAE & Beam Search \\
GLEN~\citep{Lee2023GLENGR} & NQ320k & T5-220M & Keywords & Beam Search\\
TIGER~\citep{Rajput2023RecommenderSW}  &  OpenInstIR  & Llama-3B & RQ-VAE & Reverse-Annealing \\
ZeroGR (Ours) & OpenInstIR  & Llama-3B & Title & Reverse-Annealing\\
\bottomrule
\end{tabular}
\end{table}

\begin{table}[t]
\centering\small
\caption{\edit{Performance comparison between doc2query and our method for the RQ-VAE docID baseline (TIGER~\citep{Rajput2023RecommenderSW}).}}
\label{tab:ablate-q2d}
\begin{tabular}{lccc}
\toprule
Task & doc2query & Our query generator & Diff \\
\midrule
AILA2019-Case & 2.00 & 2.00 & +0.00 \\
Apple & 13.70 & 5.48 & -8.22 \\
ArguAna & 12.00 & 11.00 & -1.00 \\
BillSum & 36.00 & 66.00 & \textbf{+30.00} \\
ClinicalTrials\_2021 & 4.67 & 6.67 & +2.00 \\
ClinicalTrials\_2023 & 1.35 & 2.70 & +1.35 \\
CodeEditSearch & 13.00 & 22.00 & +9.00 \\
CodeSearchNet & 33.00 & 56.00 & \textbf{+23.00} \\
Competition-Math & 40.00 & 61.00 & \textbf{+21.00} \\
Conala & 3.00 & 9.00 & +6.00 \\
ConvFinQA & 22.92 & 37.50 & +14.58 \\
FiQA & 7.00 & 13.00 & +6.00 \\
FinQA & 14.44 & 41.11 & \textbf{+26.67} \\
LeetCode & 6.00 & 30.00 & \textbf{+24.00} \\
LegalQuAD & 10.00 & 4.00 & -6.00 \\
LitSearch & 12.00 & 31.00 & \textbf{+19.00} \\
NFCorpus & 41.00 & 6.50 & -34.50 \\
News21 & 13.67 & 21.88 & +8.20 \\
SciDocs & 16.00 & 14.00 & -2.00 \\
SciFact & 34.00 & 42.00 & +8.00 \\
StackMathQA & 13.00 & 26.00 & +13.00 \\
TAT-DQA & 7.14 & 27.14 & \textbf{+20.00} \\
ToT\_2023 & 3.00 & 0.00 & -3.00 \\
TopiOCQA & 18.00 & 8.00 & -10.00 \\
Touche & 46.94 & 39.80 & -7.14 \\
\midrule
Average & \textbf{16.95} & \textbf{23.35} & \textbf{+6.40} \\
\bottomrule
\end{tabular}
\end{table}

\begin{algorithm}[t]
\caption{Reverse-Annealed DocID Generation}
\label{alg:reverse_annealing}
\begin{algorithmic}[1]
\REQUIRE Prefix tree $T$, decoder $f(\cdot)$, query $q$, number of docids $K$, maximum temperature $T_{\max}$
\ENSURE Generated docids $\mathcal{Z}$

\STATE $\mathcal{Z} \leftarrow \emptyset$
\FOR{$i = 1, \ldots, K$}
    \STATE Compute temperature $t_i \leftarrow g(i)$ (Eq.~\ref{eq:temperature})
    \STATE Initialize prefix $\mathbf{x}_i \leftarrow \emptyset$
    \WHILE{$\mathbf{x}_i$ is not a leaf of $T$}
        \STATE Get logits $\boldsymbol{\ell} \leftarrow f(q, \mathbf{x}_i)$
        \STATE Sample $x \sim \operatorname{Softmax}(\boldsymbol{\ell}/t_i)\big|_{T}$
        \STATE $\mathbf{x}_i \leftarrow \mathbf{x}_i \cup \{x\}$
    \ENDWHILE
    \STATE Add docid $z_i$ to $\mathcal{Z}$ and remove its leaf from $T$
\ENDFOR
\RETURN $\mathcal{Z}$
\end{algorithmic}
\end{algorithm}

\begin{figure}[t]
\begin{mdframed}
\begin{lstlisting}
def normalized_sigmoid(t, k=10, m=0.5):
    sigmoid = lambda z: 1 / (1 + np.exp(-z))
    a = sigmoid(k * (0 - m))
    b = sigmoid(k * (1 - m))
    return (sigmoid(k * (t - m)) - a) / (b - a)
\end{lstlisting}
\end{mdframed}
\caption{\edit{Normalized sigmoid function. $t$ is the step number.}}
\label{fig:normalized-sigmoid}
\end{figure}

\end{document}